\documentclass[preprint]{revtex4}

\usepackage{graphicx,amssymb}
\usepackage{dcolumn}%
\usepackage{bm}%
\usepackage{multirow}

\begin{document}

\title{About the tolerance factor for pyrochlores and related structures}

\author{R.Mouta, R. X. Silva}
\affiliation{Departamento de F\'{\i}sica, CCET, Universidade Federal do Maranh\~ao, 65085-580, S\~ao Lu\'{\i}s - MA, Brazil}
\author{C. W. A. Paschoal}
\affiliation{Curso de Ci\^encias Naturais, Universidade Federal do Maranh\~ao, Campus VII, 65400-000, Cod\'o - MA, Brazil}
\affiliation{Department of Materials Science and Engineering, University of California Berkeley, 94720-1760, Berkeley - CA, United States}
\affiliation{Department of Physics, University of California Berkeley, 94720-7300, Berkeley - CA, United States}
\email{paschoal@ufma.br}
\thanks{Permanent Address: DEFIS - UFMA - Phone: +55 98 3272 9209; Fax: +55 98 3272 8204; Alternative e-mails: paschoal.william@gmail.com; paschoal.william@berkeley.edu}


\begin{abstract}
In this work a new empirical tolerance factor for compounds with pyrochlore structure is proposed. This suggested tolerance factor is based on experimental structural data and on the tolerance factors currently proposed. However, since it does not depend on the structural data, this new tolerance factor permits the prediction of some  properties of these compounds directly. Also, a good structure stability field for the pyrochlore formation is observed when this tolerance factor is used.
\end{abstract}

\maketitle



\section{Introduction}

The tolerance factor $t$ was proposed by Goldschmidt to describe the stability and distortion in perovskite structures \cite{<spanclasshitHilite>Goldschmidt}. This geometrical parameter was defined for ABO$_3$ oxide perovskite compounds in terms of the ionic radii as:
\begin{equation}
t = \frac{R_A+R_O}{\sqrt{2}(R_B+R_O)},
\end{equation}
where $R_A$, $R_B$ and $R_O$ are the ionic radii of the A cation, B cation and oxygen, respectively. By definition, in perovskites, the tolerance factor provides a measure of how well the A-site cation fits the twelve-fold coordinated space within the corner-shared octahedral network formed by the B-site cation. Thus, this parameter indicates how far from ideal packing the ionic sizes from A and B cations can change, with fixed size for the oxygen anion, and the structure still remain an ideal perovskite. The value $t = 1$ indicates the ideal, in which the ions considered as perfect spheres, are connected in a perfect cubic lattice. Values different from the ideal indicate distortions in the structure with relation to the ideal perovskite, that are usually assumed as: (i) For $ t > 1$ , the	A cations are too large to fit into their sites and in this case hexagonal perovskites occur more frequently; (ii) for $\frac{\sqrt{2}}{2} < t < 0.9$ the A cations are too small to fit into their sites and several possible perovskite-related distorted structures are proposed such as orthorhombic, tetragonal, monoclinic and rhombohedral structures, usually originating from BX$_6$ octahedral tilting to accommodate the small A cation; (iii) for values lower than $t = \frac{\sqrt{2}}{2}$,	when A and B cations have the same size, close-packed structures are observed, as corundum, ilmenite and KNbO$_3$ type \cite{Ishihara2009a}. Although the tolerance factor is a simple geometrical parameter  based on the assumption of rigid spheres, it is a powerful tool  to predict distortions in perovskite compounds and help to propose models for their physical properties. Useful correlations have been found between $t$ and device-oriented properties of the materials. For example, Z\"urmuhlen {\it et al.} \cite{ZURMUHLEN1995} suggested that the restoring force constant of the lowest polar mode is strongly dependent on the tolerance factor, as well as the binding energy and dielectric constant of perovskites oxides. The tolerance factor is correlated to the typical highest wavenumber phonon with symmetry $A_{1g}$ active in Raman scattering \cite{Petzelt1992,ZURMUHLEN1994,ZURMUHLEN1995}.  Consequently, as the intrinsic losses are due to phonons, they define the dielectric applicability in the microwave frequency region. Even in other perovskite-related structures, such as piroxene \cite{Cheng2013} and anti-perovskites \cite{Zhao2012}, the tolerance factor plays an important role.

The oxide pyrochlores whose stoichiometry is A$_2$B$_2$O$_6$X (where the anion X can be O, F or OH) are ternary  or quaternary structures that  like the perovskites are multifunctional, serving as an appropriate crystallographic hosts for a wide range of applications, due to the great variety of possible site substitutions on both A and B sites. These substitutions imply an assortment of different physical properties, the main being magnetic frustration \cite{Gardner2010}, but several other important properties, such as catalytic \cite{Playford2011,Oh2012,Vega2012a}, topological Hall effect \cite{Ueda2012a}, metal-insulator transitions \cite{Yamaura2012a,Fujimoto2002a}, giant magnetoresistance \cite{Shimakawa1996c,Subramanian1996c}, analogous Dirac string and magnetic monopoles \cite{Morris2009a,Castelnovo2008a}, magnon hall effects\cite{Onose2010a}, metallic "ferroelectricity" \cite{Sergienko2004a}, ionic conduction \cite{Lian2001a}, superconductivity \cite{Hanawa2001d,Vyaselev2002a,Kasahara2006h},  ferroelectricity \cite{Dong2010,Dong2009} and quantum paraelectric behavior \cite{Kamba2007}, are also observed. Thus, defining a tolerance factor for compounds with pyrochlore-based structure is very useful, because it permits us  (as in the perovskite case) to predict properties before synthesizing these compounds as in the perovskite case. Accordingly Isupov \cite{Isupov1958} and Cai {\it et al.} \cite{Cai2011c} proposed different tolerance factors to describe the pyrochlore crystal structure. Isupov's tolerance factor was more elegant and resembled that used to describe perovskites, although it was not so good for describing the structure stability field for pyrochlores. This was emphasized by Cai {\it et al.} \cite{Cai2011c}, who proposed two other expressions. However, both expressions depend either on the cubic lattice parameter or on independent oxygen position parameters, which makes both difficult to calculate. In this work we derive an empirical tolerance factor for compounds with pyrochlore structure based only on the ionic radii of the constituent ions and investigate its correlation to structural, dielectric, and phonon properties.

\section{Previous tolerance factors for pyrochlores}

The first tolerance factor proposed for oxide pyrochlores was defined by Isupov \cite{Isupov1958} as
\begin{equation}
t= 0.866\frac{R_A+R_O}{R_B+R_O}.
\end{equation}
To derive it, Isupov considered the BO$_6$ octahedra as perfect. Observe that this tolerance factor is similar to that proposed by Goldschmidt to describe the stability and distortion in perovskite structures \cite{<spanclasshitHilite>Goldschmidt}. However, in the ideal pyrochlore structure A$_2$B$_2$O$_6$X, whose symmetry belongs to the space group $Fd\overline{3}m$, considering the B cation at the origin (origin 2), the A cations and the anion X  are in special positions (A is in the $16d$ Wyckoff site with coordinates $\left( \frac{1}{2},\frac{1}{2},\frac{1}{2} \right)$ and the anion X is in the $8b$ site with coordinates $\left( \frac{3}{8},\frac{3}{8},\frac{3}{8} \right)$) and the remaining oxygens are located in the $48f$ Wyckoff site with coordinates $\left(x,\frac{1}{8},\frac{1}{8} \right)$. As showed by Subramanian \cite{Subramanian1983a}, in this structure the A cation has 8-fold coordination into a scalenohedron (distorted cube) and the B cation has 6-fold coordination into a trigonal antiprism (distorted octahedron). The oxygen independent parameter $x$ defines the distortion of the A and B polyhedra coordination. Geometrically, when $x=0.3125$, the BO$_6$ polyhedron is a regular octahedron while AO$_8$  polyhedron is a distorted cube (scalenohedron), while for $x=0.375$, the AO$_8$  polyhedron is an ideal cube and BO$_6$ is a distorted octahedra. Thus, the octahedron and the cube cannot both be regular. Furthermore, the limiting value for these compounds to become fluorites is $x>0.375$. Usually, the $x$ parameter lies between 0.309 and 0.355 and these values imply distorted cubes and octahedra simultaneously \cite{Subramanian1983a}. Therefore, the Isupov's assumption is hardly achieved in pyrochlore structure leading to incoherent tolerance factor calculations.

To consider the effect of essential structural features on the tolerance factor, Cai {\it et al.} \cite{Cai2011c} recently proposed two tolerance factors to describe the pyrochlore crystal structure. Cai {\it et al.} considered the geometrical features of the different cation coordination polyhedra and proposed two distinct tolerance factors:
\begin{equation}
t_1 = \frac{\sqrt{\left(x-\frac{1}{4}\right)^2 +\frac{1}{32}} }{\sqrt{\left(x-\frac{1}{2}\right)^2 +\frac{1}{32}}} \frac{R_A+R_O}{R_B+R_O}
\end{equation}
and
\begin{equation}
t_2 = a \frac{\sqrt{3}}{8(R_A+R_O)}.
\end{equation}

The $t_1$ and $t_2$  parameters proposed by Cai {\it et al.} are tolerance factors corresponding to A$_2$B$_2$ and XA$_4$ polyhedra, respectively. In the first one the O anion is present, while in the second, the X anion. Mainly based on the $t_1$ parameter, Cai {\it et al.} proposed a stability field to distinguish pyrochlores from weberites (fluorite-related structures). A relationship between the tolerance factors and dielectric properties and their applicability to analyze structure-property relations was discussed. However, both defined tolerance factors depend either on the cubic lattice parameter or on independent oxygen position parameters. The introduction of these structural parameters makes the tolerance factors proposed by Cai {\it et al.} more precise, but this complicates their calculations, because, as pointed out by Cai {\it et al.} it is necessary to measure $a$ and $x$ data to their calculation. Although we can use theoretical estimates based on the Nikiforov \cite{Nikiforov1972} and bond valence sum methods \cite{Brown2006} for these parameters,  this greatly limits the structural predictions for pyrochlore compounds. Therefore, an empirical tolerance factor that does not depend on the structural parameters, just on the ionic radii, is proposed in this work.

\section{New empirical tolerance factor for pyrochlore structure}
As discussed, the main goal is to obtain an expression for the tolerance factor for pyrochlores similar to that proposed by Goldschmidt for perovskites, which depends only on the ionic radii of the constituent ions. For this, we started from the expressions proposed by Cai et al. \cite{Cai2011c}. Therefore, we need to obtain either $a$ or $x$ as a function only of the ionic radii, in a manner that fits well into the available experimental data at room temperature and atmospheric pressure. However, to model accurately the $x$ parameter, one would need to employ mainly neutron diffraction data, due to the well-known difficulty of determining precise coordinates of light atoms, such as oxygen, by X-ray diffraction. Although this care avoids misleading results, it also reduces greatly the data amount that can be used in the fit. Thus, we started from the $t_2$ parameter and obtained an expression for the lattice constant.

Recently, Brik and Srivastava \cite{Brik2012} succeeded in predicting lattice constants for several pyrochlores. However, the expression proposed by them that depends only on the ionic radii is inadequate to model pyrochlores with mixed cations and anions. Thus, their main expression included not only ionic radii, but also electronegativities of the constituting ions, which is, again, undesirable for our purposes. So, in this work we tested many forms of dependence of the lattice constant with $R_A + R_O$ and $R_B + R_O$ for the same ternary pyrochlores used in Brik and Srivastava's work, being 79 distinct compounds in total\footnote{We excluded theoretical values and measured values that were not taken at room temperature and atmospheric pressure.}. In those cases of materials whose lattice parameters have more than one experimental value  reported, all of them were considered, which leads to 110 different values. All the experimental data used are available as supplementary material, along with the corresponding ICSD (Inorganic Crystal Structure Database) \cite{icsd1,Belsky2002} reference for each lattice constant value.

\begin{figure}
  \centering
  \includegraphics[width=\textwidth]{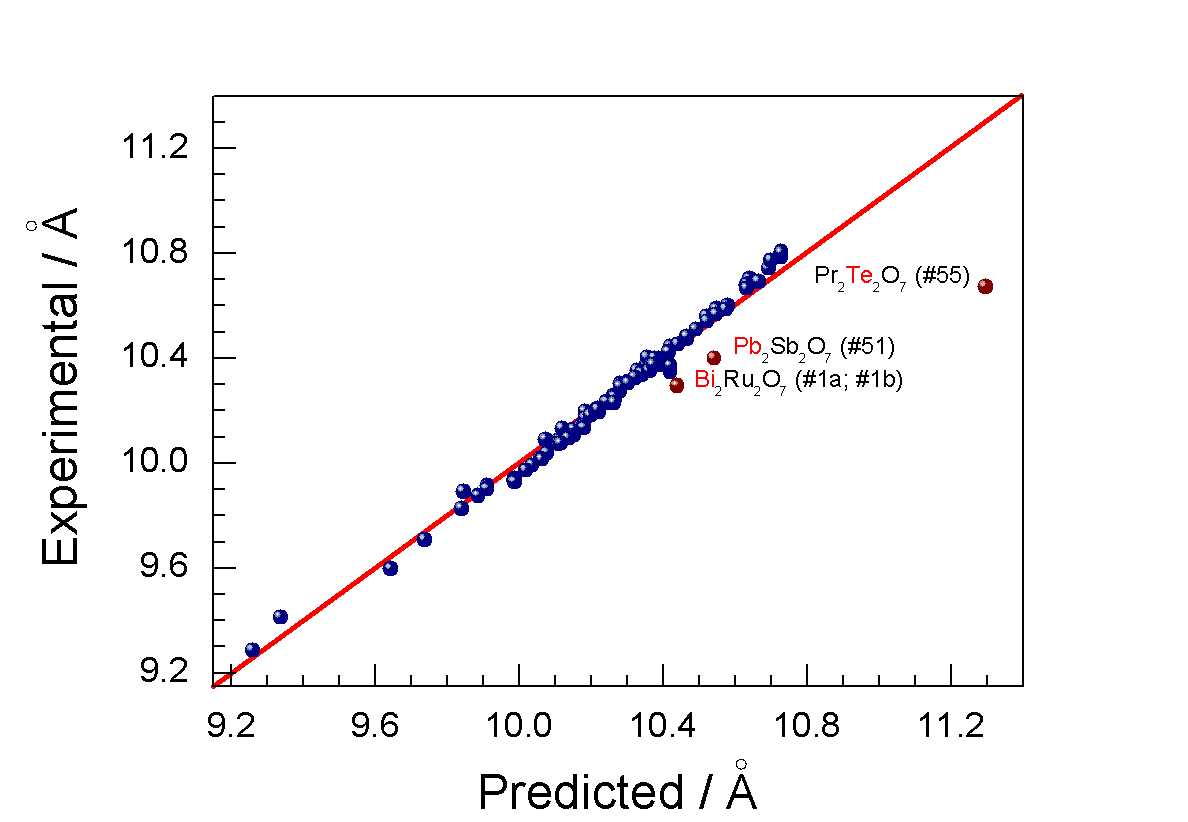}\\
  \caption{Comparison between the experimental and calculated lattice constants for pyrochlores. The red line is a guide for the eyes, representing the ideal case where the experimental and calculated values perfectly match. The red dark data points (labeled) correspond to the compounds containing an ion with a stereochemically active non-binding electron lone pair. The labels in parenthesis refer to Tables 1, 2 and 3 of the supplementary material.  }\label{figure1}
\end{figure}

\begin{figure}
  \centering
  \includegraphics[width=\textwidth]{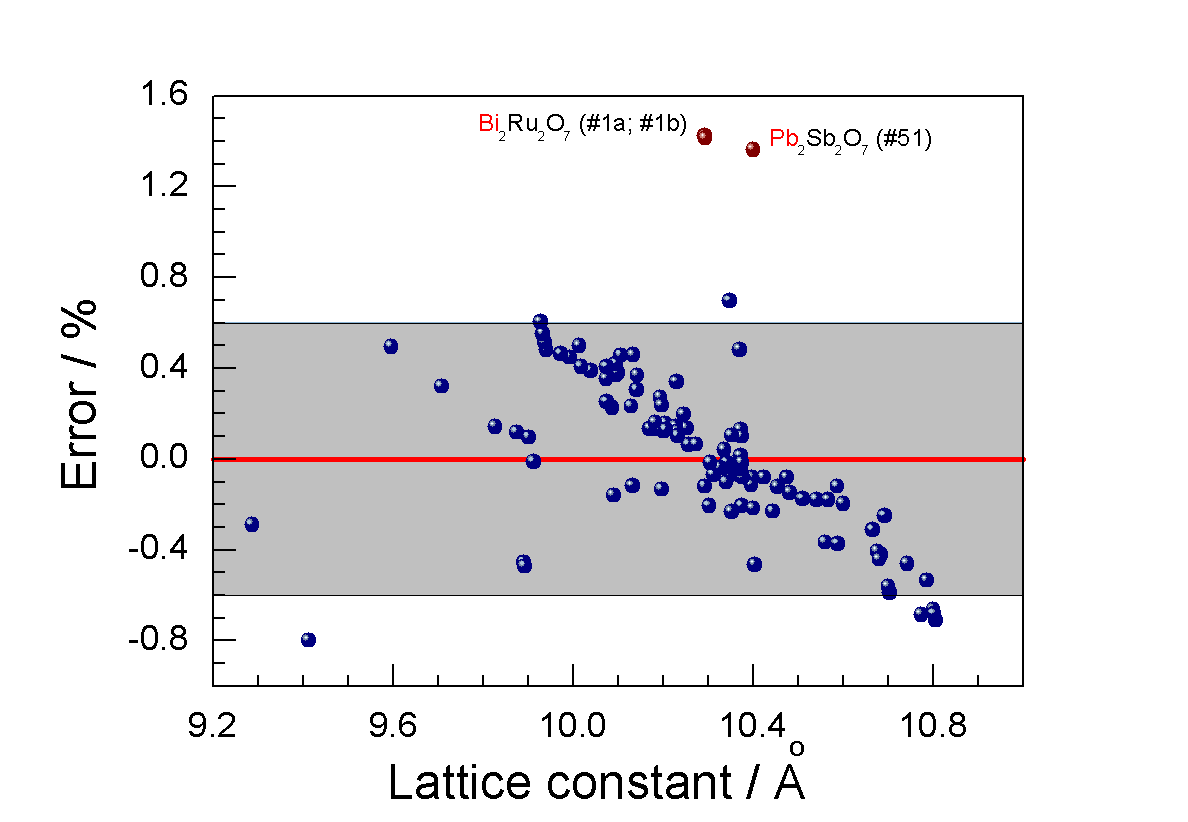}\\
  \caption{Percent errors between the calculated and experimental lattice constants. The light blue region encloses compounds for which the percent deviation is not greater than 0.6 \%. The red dark data points have the same meaning as in Figure 1. Since the error for Pr$_2$Te$_2$O$_7$ was too large, it was not shown. Refer to Table 1 of the supplementary material for numerical values.  }\label{figure2}
\end{figure}

From all the expressions tested, the one that best fit the experimental data was
\begin{equation}\label{a}
  a = \frac{8}{\sqrt{3}} \left [ 1.43373(R_A+R_O )-0.42931 \frac{\left(R_A+R_O\right)^2}{R_B+R_O} \right]
\end{equation}
where we took the oxygen ionic radius to be 1.38 \AA. The comparison between the experimental and calculated lattice constants obtained is shown in Figure \ref{figure1}, while the percent errors with respect to the experimental data are shown in Figure \ref{figure2}. The data points marked in dark red correspond to compounds containing ions with a stereochemically active non-binding electron lone pair. These were not used in the fit, since in this case the concept of a spherical ion (implied by the assumption that it possesses an ionic radius) is not so good anymore. Besides, such lone pairs tend to repel the neighboring oxygen ions, distorting the structure and making it deviate from the expected structure. Accordingly, these compounds noticeably correspond to the most poorly reproduced values, as one can expect for a model based on spherical ions.

Now we compare our model for the lattice constant with the model of Brik and Srivastava\footnote{All the experimental and calculated values for the lattice constants, along with the experimental x parameters and the employed ionic radii are available as supplementary material.}. For the compounds considered here (see Table 1 in the supplementary material), both models provide a root-mean-square deviation between the calculated and experimental of about 0.02 $\AA$, but our average percent error (0.27 \%) is subtly lower than theirs (0.41 \%) \footnote{We considered the absolute values of the percent errors, in order to avoid cancelations between positive and negative values }. Except for three pyrochlores presenting lone pair effect (Pr$_2$Te$_2$O$_7$ (5.85 \%), Bi$_2$Ru$_2$O$_7$ (1.42 \%) and Pb$_2$Sb$_2$O$_7$ (1.36 \%)) the error does not exceed 1\%; for 6 compounds, the errors lie between 0.6 \% and 1 \%; all the remaining crystals have an error lower than 0.6 \%. Thus, we have a reliable prediction for the lattice parameter of the pyrochlores based uniquely on the ionic radii. The error predicting the lattice parameter by our model for the mixed pyrochlores cited by Brik and Srivastava is also subtly lower than the error they obtained. For those compounds the average percent error is 0.45\% with a maximum value of 1.82\%. Thus, we preferred to model the lattice parameter using the equation \ref{a} rather than that proposed by Brik and Srivastava \cite{Brik2012} that depends only on the ionic radii.

Finally, since we have established that equation \ref{a} provides a very good estimate of the lattice constant for pyrochlores, we propose a new empirical equation for the tolerance factor for pyrochlore oxides compounds. Using the Cai {\it et al.} expression and substituting the obtained $a$ value, we have
\begin{equation}\label{new_t}
  t = 1.43373-0.42931 \left( \frac{R_A+R_O}{R_B+R_O} \right)
\end{equation}
We stress that the advantage of this expression lies on the absence of the explicit dependence on structural parameters (although it takes them into account), which allows us to predict structural and properties features of pyrochlore compounds before measuring or modeling/estimating their structural data, as is the case for perovskites. Interestingly, the tolerance factor decreases with $\left( \frac{R_A+R_O}{R_B+R_O} \right)$, which is different from what has been proposed by Isupov for pyrochlores and by Goldsmith for perovskites. This suggests a limit value for the tolerance factor, because this can not be negative or null.

\begin{figure}
  \centering
  \includegraphics[width=\textwidth]{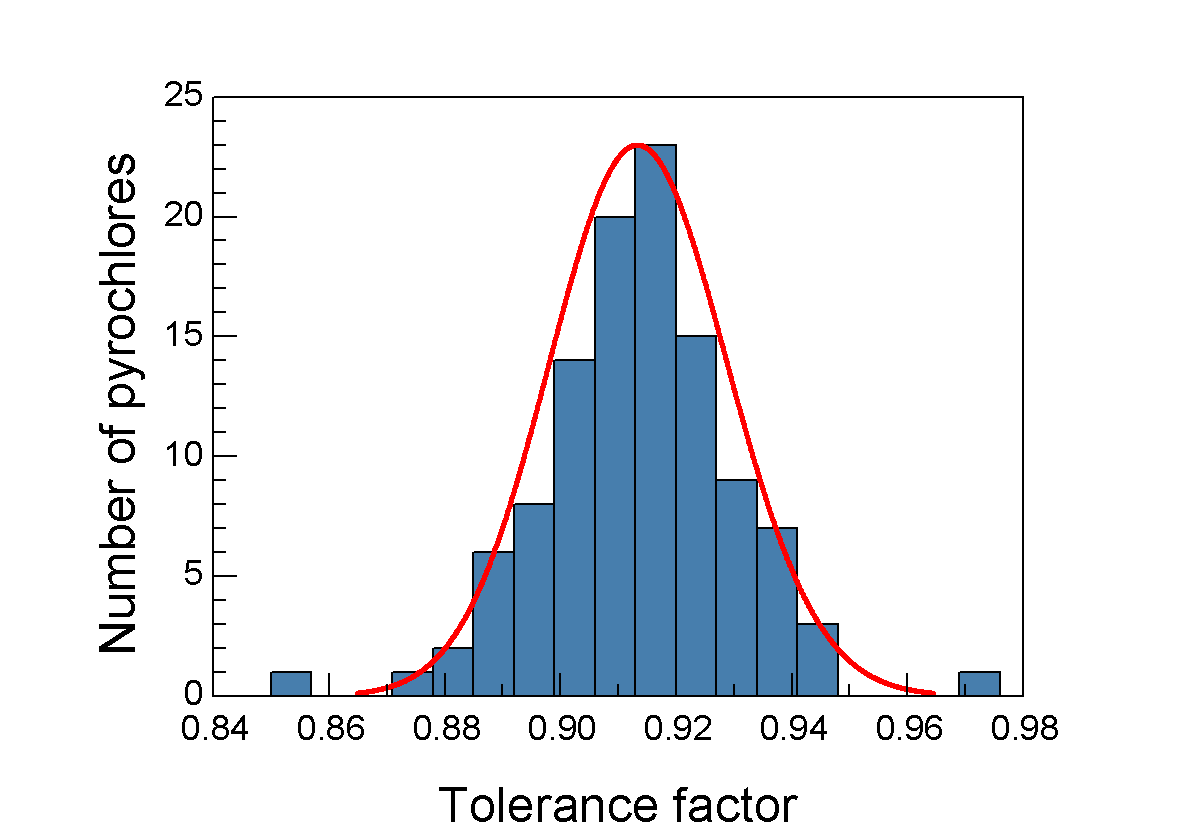}\\
  \caption{The distribution of pyrochlore compounds based on the new tolerance factor}\label{histogram}
\end{figure}

The histogram that describes the pyrochlore distribution according to the new tolerance factor is shown in Figure \ref{histogram}. We tested the distribution according to three normal distribution tests: Shapiro-Wilk, Lilliefors, and Kolmogorov-Smirnov. For all, at the level of 0.05, the data was significantly drawn from a normally distributed population. Therefore, although the new tolerance factor is based on the $t_2$ tolerance factor proposed by Cai {\it et al.}\cite{Cai2011c}, the new distribution is symmetrically distributed. Thus, we fit the histogram with a gaussian curve (shown in Figure \ref{histogram}), whose centre occurs at the tolerance factor of  $t=0.913$. For pyrochlores for which the tolerance factor is near this value, the radii ratio $R_A/R_B$ is near 1.6. This value makes sense since it is almost in the middle of the $R_A/R_B$ range (l.46 to 1.80) proposed by Subramanian {\it et al.} \cite{Subramanian1983a} as necessary for pyrochlore formation.

\section{Properties correlated to the new tolerance factor}
\begin{figure}
  \centering
  \includegraphics[width=\textwidth]{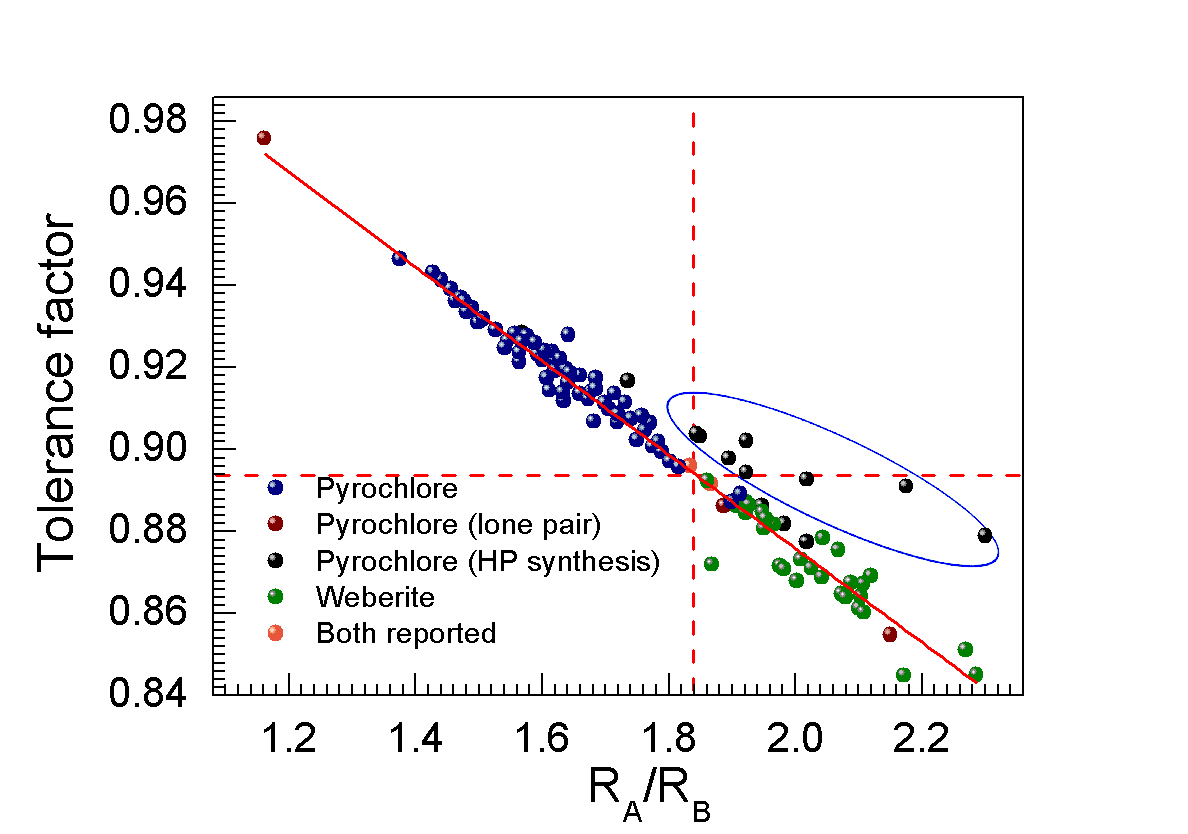}\\
  \caption{Tolerance factor as a function of the A and B cations ionic radii indicating the separation between weberite and pyrochlore structure. The vertical and horizontal red dashed lines indicate the best point to distinguish pyrochlores from weberites according to the radii ratio and tolerance factor. The red inclined line indicates a linear fit $\left( R^2 = 0.9704 \right)$, whose expression is $t=1.1045-0.1143\left( \frac{R_A}{R_B} \right)$.  }\label{stability}
\end{figure}

One of the most important application of tolerance factors is to estimate the structure stability field. Previously, Subramanian {\it et al.} \cite{Subramanian1983a} reported for pyrochlores that the A and B ionic radii could be used to define the stability field for A$_2^{3+}$B$_2^{4+}$O$_7$, which was driven by the radii ratio, $\left( \frac{R_A}{R_B} \right)$, and the independent oxygen coordinate, $x$. In their work, Cai {\it et al.} \cite{Cai2011c} used the tolerance factor $t_1$ to establish a stability field for pyrochlores, to mainly distinguish between weberites and pyrochlores structures. In this case, they obtained a good stability field, with a separation between weberites and pyrochlores. But, due to the meaningless value of $x$ for weberites, $t_1$ may not be the best way to distinguish pyrochlores and weberites according to the authors \cite{Cai2011c}. In Figure \ref{stability} we plot the stability field based on the new tolerance factor for compounds with pyrochlore and weberite structures. From our stability field it is clear we distinguish between pyrochlores and weberites based just on the tolerance factor, as clear as the radii ratio and better than the one proposed by Cai et al {\it et al.} \cite{Cai2011c}. Observe we do not have any problem with the absence of structural data for weberites because we need just the ionic radii.
The structural stability field for oxide pyrochlores was well described by Subramanian {\it et al.} using the cation radii ratio. We can see from Figure \ref{stability}, our tolerance factor defines a stability field similar to that defined by the radii ratio. As well as our tolerance factor, the radii ratio can not differentiate weberites and pyrochlores synthesized at high pressures. However, when we use both together, these parameters can predict the pyrochlores synthesized at high temperatures, as indicated in the Figure \ref{stability} (see data inside the ellipse), which are obtained, preferentially, using radii ratio for weberites and tolerance factor for pyrochlores.

\begin{figure}
  \centering
  \includegraphics[width=\textwidth]{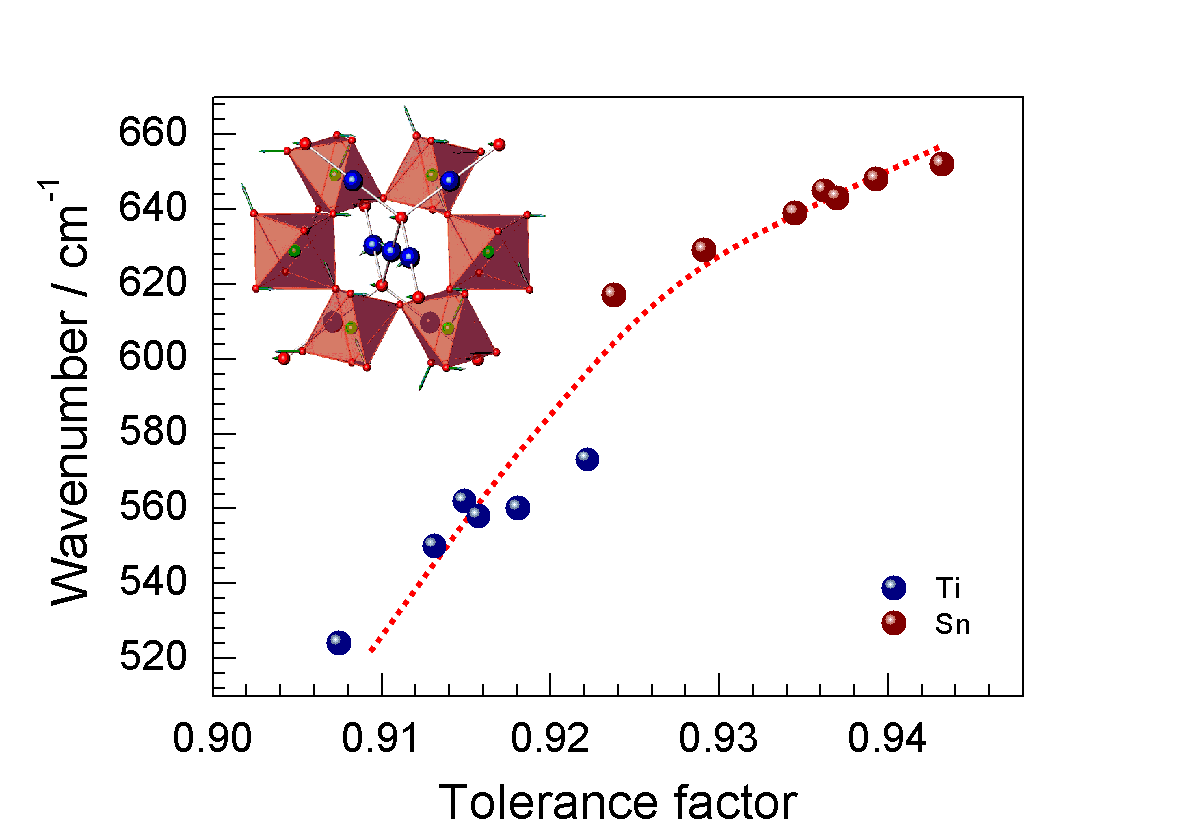}\\
  \caption{Correlation between the tolerance factor and the highest infrared-active observed mode. The red dashed line is a guide for the eye. The inset shows the phonon vibration according to force field proposed by Silva {\it et al.} \cite{silva}.}\label{F1u-7}
\end{figure}

\begin{figure}
  \centering
  \includegraphics[width=\textwidth]{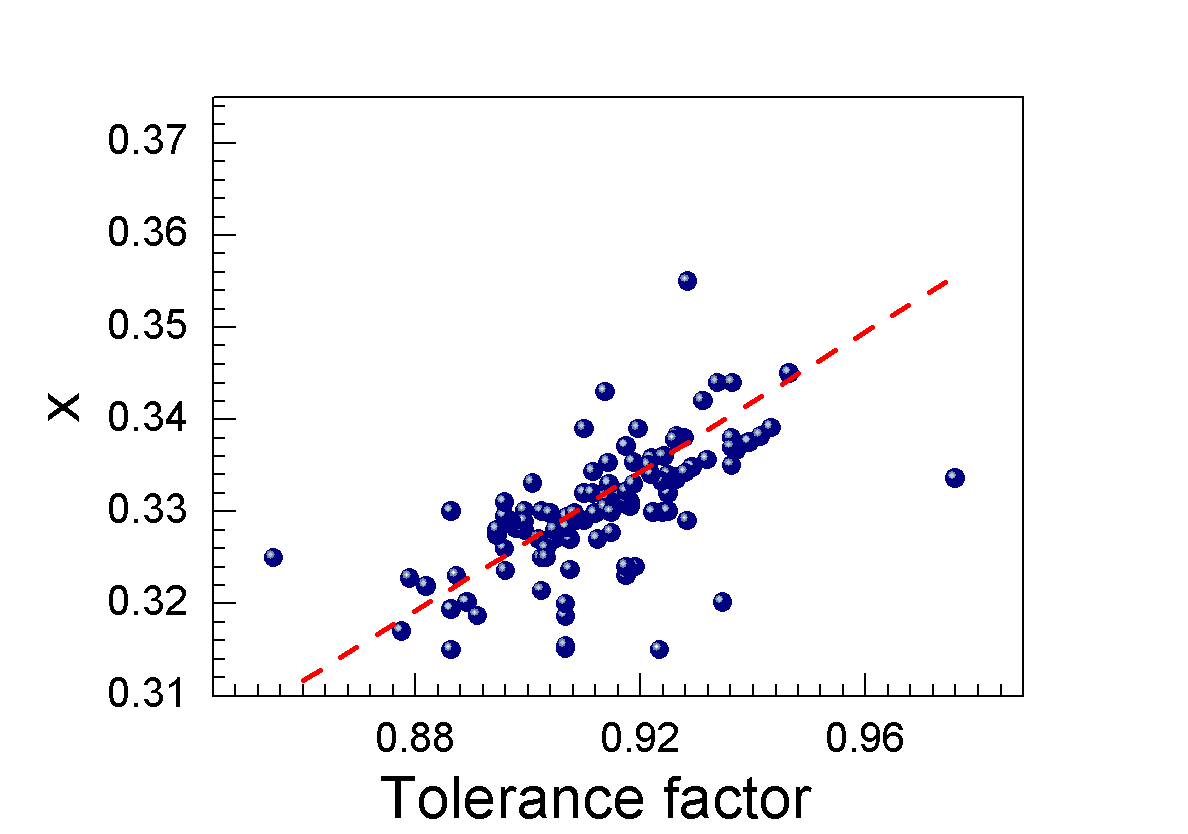}\\
  \caption{Correlation between the tolerance factor and the structural oxygen parameter $x$. The line indicates a linear fit $\left(x=-0.0147+0.3794t\right)$ to indicate the trend. The scale of the y axis was chosen to show all possible $x$ parameter values.}\label{x_t}
\end{figure}

In perovskites, the most remarkable correlations between the tolerance factor and properties are associated to the phonons and dielectric constant. Figure \ref{F1u-7} shows for pyrochlore compounds with B$=$Sn or Ti the behaviour with tolerance factor of the highest infrared-active mode experimentally observed, whose symmetry is $F_{1u}$.  Infrared-active phonons are important, because they are directly connected to the dielectric constants. The behaviour demonstrates a strong correlation with the new tolerance factor, whose Pearson product-moment correlation coefficient is $0.97$. For all remaining infrared-active phonons an ionic mass dependence was observed when correlated to the tolerance factor. The mass effect is not observed for the highest phonon because in this vibration the A and B ions are almost stopped, being the motion associated mainly with the oxygens in the pseudo-octahedra (see inset in Figure \ref{F1u-7}). The increasing behaviour can be explained considering the oxygen motion in this phonon. The motion changes linearly the B-O-B angle, which changes linearly with the oxygen parameter $x$. \cite{Subramanian1983a} But, $x$ parameter increased with the tolerance factor for the considered B ions (B = Ti and Sn), implying in this increasing behavior for the phonon under tolerance factor variation. In fact, there is a small correlation (Pearson product-moment correlation coefficient of $0.65$ when considering just parameters determined by neutron and synchrotron data) between $x$ parameter and tolerance factor, as shown in Figure \ref{x_t}. Observe that oxygen parameter $x$ exhibits a clear trend in its values as a function of tolerance factor, increasing when the tolerance factor increases.

Finally, we can show that our tolerance factor exhibits the same correlation as the one proposed by Cai {\it et al.} \cite{Cai2011c} for the normalized difference in dielectric permittivity $\Delta\varepsilon_n$, as showed in Figure \ref{dielectric}. This parameter is defined in terms of the measured  dielectric constant $\varepsilon_m$, the permittivity obtained by the Clausius-Mossotti $\varepsilon_r$ and the microscopic polarizability, $\alpha$, as:
\begin{figure}
  \centering
  \includegraphics[width=\textwidth]{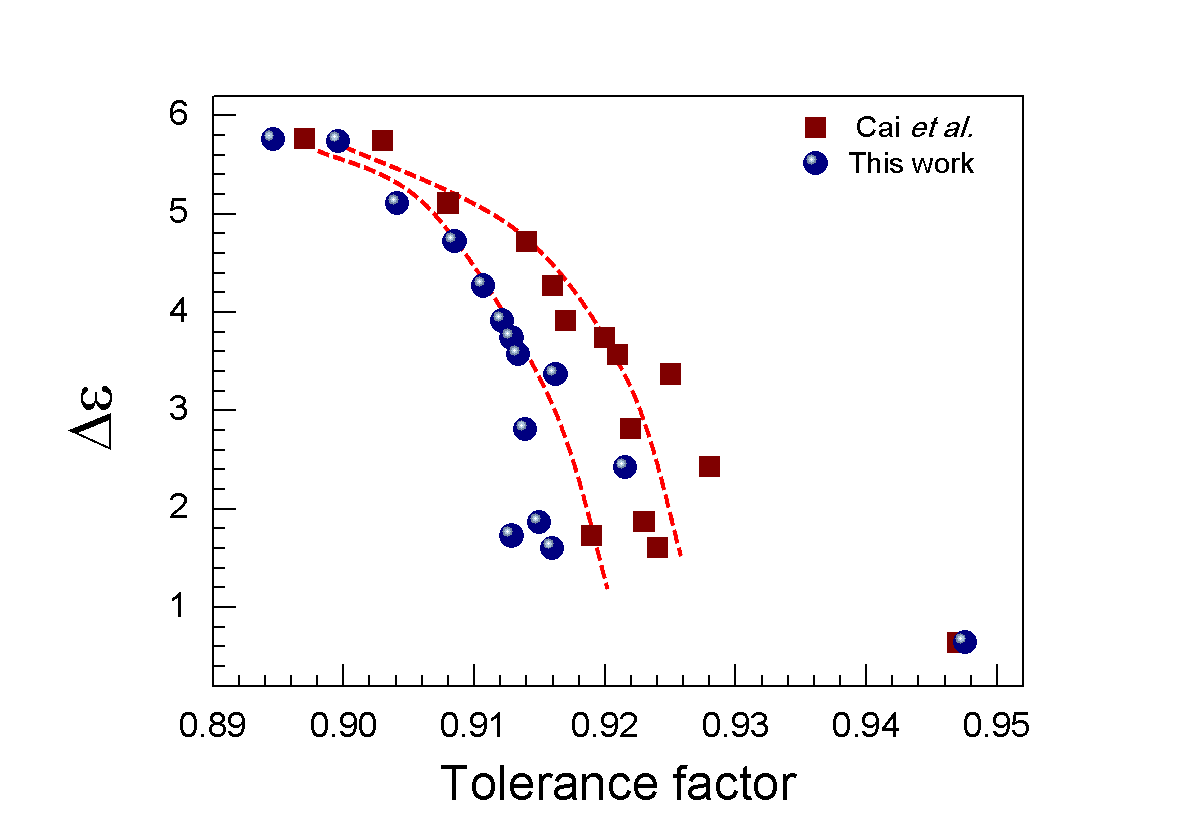}\\
  \caption{Normalized permittivity difference dependence with the tolerance factor. The values calculated by Cai {\it et al.} are also shown. The red dashed lines are a guides for the eye. The compounds used in this figure are listed in Table 4 of the supplementary material.}\label{dielectric}
\end{figure}
\begin{equation}\label{e}
  \Delta\varepsilon_n = \frac{\varepsilon_m - \varepsilon_r} {\alpha}
\end{equation}

\section{Conclusions}
In summary, we have proposed an empirical tolerance factor for compounds with pyrochlore structure based only on the ionic radii of the constituent ions. The pyrochlore distribution according to the new tolerance factor is symmetrically centered at the value $t=0.913$, which is an ideal value connected to the $R_A/R_B$ ratio proposed by Subramanian {\it et al.} as necessary for pyrochlore formation. The new tolerance factor permits a good prediction of the lattice parameter and exhibits a strong correlation with several properties, such as those related to the stability, phonons, and dielectric constants.

\section*{Acknowledgements}
This work was partially supported by the Brazilian funding agencies CAPES, CNPq, and FAPEMA. The authors are grateful to Dr. Michael Lufaso for all discussions around the theme and valuable suggestions.

\bibliography{Pyrochlores}

\end{document}